\documentclass[useAMS,usenatbib]{mn2e}

\usepackage{amsmath}
\usepackage{amssymb}
\usepackage{amsfonts}
\usepackage{graphicx}
\usepackage{array}
\usepackage{booktabs}

\begin{document}
\title{Mode Spectrum of the Electromagnetic Field in Open Universe Models}
\author[J.\ Adamek, C.\ de~Rham and R.\ Durrer]{Julian Adamek,$^1$\thanks{jadamek@physik.uni-wuerzburg.de} Claudia de~Rham$^2$\thanks{claudia.derham@unige.ch} and Ruth Durrer$^2$\thanks{ruth.durrer@unige.ch}\\
$^1$Institut f\"ur Theoretische Physik und Astrophysik, Julius-Maximilians-Universit\"at W\"urzburg,\\ \, Emil-Fischer-Str.\ 31, 97074 W\"urzburg, Germany\\
$^2$D\'epartement de Physique Th\'eorique \& Center for Astroparticle Physics, Universit\'e de Gen\`{e}ve,\\ \, 24 Quai Ernest Ansermet, 1211 Gen\`{e}ve 4, Switzerland}

\maketitle

\begin{abstract}
We examine the mode functions of the electromagnetic field on spherically symmetric backgrounds with special
attention to the subclass which allows for a foliation as open Friedmann-Lema\^\i tre (FL) spacetime. It
is well-known that in certain scalar field theories on open FL background there can exist so-called supercurvature
modes, their existence depending on pa\-ra\-meters of the theory. Looking at specific open universe models,
such as open inflation and the Milne Universe,
we find that no supercurvature modes are present in the spectrum of the electromagnetic field. This excludes the
possibility for superadiabatic evolution of cosmological magnetic fields within these models without relying on new physics or breaking the conformal invariance of electromagnetism.
\end{abstract}

\begin{keywords}
 cosmology: theory -- magnetic fields -- early Universe.
\end{keywords}


\section{Introduction}
The generation of large scale coherent magnetic fields in the Universe which are observed
in low and high redshift galaxies~\citep{Kronberg:1993vk,Pentericci:2000mp},
clusters~\citep{Clarke:2000bz}, filaments~\citep{Battaglia:2008ex}
and even in voids~\citep{Neronov:2010,Taylor:2011bn}, is still an unsolved problem
in cosmology. Fields generated in the early Universe are generally small and, on large scales,
they usually simply evolve via flux conservation, $B\propto a^{-2}$, where $a$ denotes the
scale factor of cosmic expansion. One exception to this rule are helical magnetic fields
which develop an inverse cascade moving power from small to larger
scales~\citep{Banerjee:2004df,Campanelli:2007tc}. This can alleviate the problem of magnetic
field generation somewhat but is still not sufficient~\citep{Caprini:2009pr,Durrer:2010mq}.

Another idea has been put forward recently in \citet{Barrow:2011ic}: in an open universe, supercurvature
modes decay slower that $1/a^2$ and can therefore remain relevant at late times. The question remains
of how such supercurvature modes are generated.
In this paper, we explore this proposal within explicit open universe models.
Whilst open inflation is no longer the most favored model of inflation, it is the most explicit model that leads to an open universe, and we therefore
start by studying the generation of supercurvature modes within that model. We show that within the Coleman-de~Luccia bubble
universe~\citep{Coleman:1980aw,Bucher:1994gb}, supercurvature modes of the magnetic field are actually not part of the physical spectrum
and can therefore not be generated. We show that the same results also hold for the Milne Universe.
These are two explicit cases without Big Bang singularity. For them we can unambiguously specify the initial
Cauchy surface needed to define the quantum vacuum and the physical spectrum. Of course we could arbitrarily
pronounce any open Friedmann slicing $\{t= {\rm const.}\}$ as our initial Cauchy surface. But besides the fact
that this surface does not allow supercurvature modes, such a definition is arbitrary, incomplete and not unique.

Let us first, in  a brief paragraph, present the issue of supercurvature modes.
The Laplacian on the spatial slices $\{ t=\mbox{const.}\}$ in an open Friedmann universe with
curvature $K$ has eigenfunctions with eigenvalues $-k^2$,
$$ \Delta Y_k = -k^2 Y_k \,.$$
The functions with $k^2>|K|$ or, for symmetric, traceless tensors of rank $m$,  with $k^2>(1+m)|K|$,
form a complete set of functions
on these slices which reduces to the usual Fourier modes in the limit $K\rightarrow 0$. There are,
however, also so-called supercurvature modes, eigenfunctions of the Laplacian with
eigenvalues in the range $0<k^2<|K|$. At first
glance one might argue that, since every square integrable
function can be expanded in terms of the subcurvature basis, supercurvature
modes play no role. If we only consider the post-inflationary Universe, this view seems justified.
However, it has been shown in \citet{Sasaki:1994yt}, that in open inflation under certain circumstances supercurvature modes can be present, see also \citet{Lyth:1995cw} and \citet{GarciaBellido:1995wz} for a discussion in a different context. The basic reason for this is that the $\{ t=\mbox{const.}\}$
slices of the post-inflationary Universe do not represent Cauchy surfaces of the entire spacetime containing the Coleman-de~Luccia
bubble. However, in order to discuss quantum fields and particle
generation during inflation,
we have to expand the fields in a complete basis on a Cauchy surface of the inflationary Universe
and, as has been shown in~\citet{Sasaki:1994yt}, in certain cases this can lead to the generation of modes
which correspond to supercurvature modes after inflation.

In \citet{Sasaki:1994yt} the analysis is presented for scalar fields. In this paper we reduce the case
of the electromagnetic field to the scalar field problem so that we can apply the results of~\citet{Sasaki:1994yt}.
We show that when expressing the electromagnetic field in terms of the Debye potentials~\citep{TseChin:1973mp}, these
can be viewed as two conformally coupled fields for which no supercurvature modes
exist. A definite statement can be made in any setup which allows for the Cauchy problem to be
well-posed. On physical grounds, although there may be no unique mechanism to procure an open FL universe, we want to focus on the open inflation scenario. We think that a case study within this
scenario is most useful because it is by far the most
explicit and physically
well-motivated setup which naturally leads to an open FL universe and, at the same time, comes with a complete
description in the Cauchy sense. Other scenarios which, e.g., impose an open geometry to be realized ad hoc have
to be supplemented by some arbitrary assumptions, and the question of supercurvature modes can therefore not
seriously be addressed.

The open inflation scenario \citep{Bucher:1994gb,GarciaBellido:1997te} was originally introduced at a time when
observational data seemed to favor an open universe and it was therefore imperative to look for appropriate models.
With the advent of precision measurements of the anisotropies in the cosmic microwave background, the evidence for considerable curvature to be present in our Universe has virtually evaporated
\citep{Jaffe:2000tx,Spergel:2003cb}. However, the scenario has recently attracted new interest in the context of eternal
inflation \citep{Linde:1986fd,Guth:2007ng} and the Landscape idea \citep{CarrWeinberg,CarrSusskind}. From this new
point of view,
open inflation in the Coleman-de~Luccia bubble Universe remains conceptually well-motivated, although the focus has
shifted away from procuring non-vanishing curvature. In fact, the scenario allows that the curvature we observe today
can be rather minuscule, see e.g.\ \citet{DeSimone:2009dq} for a discussion. As already pointed out, in this paper
we exploit the fact that the setting contains enough information about the background spacetime such that
the questions we want to study can be addressed in a meaningful way.
Here we are not so much interested in the question of whether the spatial curvature of the observed
Universe is negative, but we want to analyze the conceptual question of whether an open universe can allow for
supercurvature modes of the electromagnetic field.

The remainder of this paper is organized as follows: in the next section we introduce the Debye potentials, write the electromagnetic Lagrangian in an open, closed or flat FL universe in these variables and derive a complete set of solutions to the Euler-Lagrange equations. We then analyze whether supercurvature modes are normalizable on a Cauchy surface, and by comparison with the pure scalar case, we conclude that no supercurvature modes are normalizable on an open de Sitter geometry. By conformal invariance we find that the same result holds for a Coleman-de~Luccia bubble. Finally, for completeness, we provide the full quantization prescription of the Debye potentials, before summarizing our results. In an appendix, we present the explicit computation of the mode normalization also for the Milne model,
which is one of the simplest open universe models.
\vspace{-2pt}


\section{General Formalism}

We consider background geometries of the Friedmann-Lema\^\i tre (FL) type. The line element reads
\begin{equation}
\label{eq:metric}
ds^2~=~-dt^2 + a^2(t) \left[dr^2 + s^2(r) d\Omega^2\right]~,
\end{equation}
where $s(r) = \sin r, r, \sinh r$ corresponds to closed, flat and open spatial geometry, respectively. In
this work, we will focus on the latter two cases. In particular, the case $a(t) \equiv \mathrm{const.}, s(r) \equiv r$ gives
the Minkowski metric, while $a(t) \equiv \sinh(H t) / H, s(r) \equiv \sinh r$ represents an open foliation of
de~Sitter space with $\Lambda = 3 H^2$. Note that with this convention, $r$ and $s$ have no units and spatial curvature is $K=\pm1$ or $0$, but
$a$ and $t$ have units of length. As usual, we set $c = \hbar = 1$.

Because these types of backgrounds are spherically symmetric, they are appropriate for studying the
electromagnetic field in terms of the Debye potentials \citep{TseChin:1973mp}. In this formalism, instead of making
use of the usual $A^\mu$ vector potential, the electromagnetic field is decomposed into two potentials $U$ and $V$.
The advantage of these Debye potentials is the fact that the equations completely decouple in any spherically
symmetric background, while the components of $A^\mu$ are usually badly mixed if the spacetime is not flat. Therefore,
the Debye potentials allow for even more general metrics than (\ref{eq:metric}).

In equations ($4$) and ($5$) of \citet{TseChin:1973mp}, expressions for the physical electric and magnetic fields
are given in terms of the Debye potentials. It will be useful for us to look at these fields in the helicity basis.
Given an orthonormal basis $(\mathbf{e}_\theta, \mathbf{e}_\phi)$ on the sphere, the helicity basis reads
\begin{equation}
\mathbf{e}_+ = \frac{1}{\sqrt{2}} \left(\mathbf{e}_\theta - i \mathbf{e}_\phi\right)~,\qquad
\mathbf{e}_- = \frac{1}{\sqrt{2}} \left(\mathbf{e}_\theta + i \mathbf{e}_\phi\right)~.
\end{equation}
We find the following components of the physical electric and magnetic field in this new basis:
\begin{eqnarray}
\label{eq:EB}
E_r &=& -\frac{1}{2 a s} \left(\eth\eth^\ast + \eth^\ast\eth\right) V~,\nonumber\\ B_r &=& -\frac{1}{2 a s} \left(\eth\eth^\ast + \eth^\ast\eth\right) U~,\nonumber\\
E_+ &=& -\frac{1}{\sqrt{2} a s} \left[\partial_r \left(s \eth V\right) + i \partial_t \left(a s \eth U\right)\right]~,\nonumber\\
B_+ &=& -\frac{1}{\sqrt{2} a s} \left[\partial_r \left(s \eth U\right) - i \partial_t \left(a s \eth V\right)\right]~,\nonumber\\
E_- &=& E_+^\ast~,\qquad B_- = B_+^\ast~.
\end{eqnarray}
In these expressions, we make use of the \textit{spin-raising}
and \textit{spin-lowering} operators $\eth$ and $\eth^\ast$. These are defined as
\begin{eqnarray}
\eth \chi &=& -\sin^\sigma\!\theta \partial_\theta \left(\sin^{-\sigma}\!\theta \chi\right) - \frac{i}{\sin \theta} \partial_\phi \chi\nonumber\\
\eth^\ast \chi &=& -\sin^{-\sigma}\!\theta \partial_\theta \left(\sin^{\sigma}\!\theta \chi\right) + \frac{i}{\sin \theta} \partial_\phi \chi~,
\end{eqnarray}
where $\sigma$ is the spin-weight of the field $\chi$. As the names imply, the spin-raising and lowering operators
increase or decrease the spin-weight of a field by one unit. See, e.g., \citet{Goldberg:1966uu} for some details
concerning these operators and the spherical harmonics used to expand a field of arbitrary spin-weight.

Using $F_{\mu\nu} F^{\mu\nu} = 2 \mathbf{B}^2 - 2 \mathbf{E}^2$, the Maxwell action in terms of the Debye potentials is
\begin{multline}
\label{eq:action}
\mathcal{S}_\mathrm{em}~=~\frac{1}{2} \int\! a^3 dt s^2 dr d\Omega \Biggl[- \frac{1}{a^2}
\partial_t \left(a \eth U\right) \partial_t \left(a \eth^\ast U\right) \Biggr.\\\Biggl.+ \frac{1}{a^2 s^2}
\left(\eth^\ast \eth U\right) \left(\eth \eth^\ast U\right) + \frac{1}{a^2 s^2}
\partial_r \left(s \eth U\right) \partial_r \left(s \eth^\ast U\right)\Biggr]\\ - \left\lbrace U \rightarrow V\right\rbrace~.
\end{multline}
It is evident from the action
that the physical degrees of freedom are carried by the helicity-one representations $\eth U$ and $\eth V$. This is
in agreement with the well-known properties of a photon. Furthermore, it is not surprising that the two helicity degrees
of freedom decouple in a spherically symmetric background.

The equations of motion are\footnote{As in
eqs.~($6$) and ($7$) of \citet{TseChin:1973mp},
we have omitted an overall spherical Laplacian $(\eth \eth^\ast + \eth^\ast \eth) / 2$ acting on the equation.
The solutions are identical up to modes which are
annihilated by this operator. These are exactly the modes of $U$ with zero angular momentum ($\ell = 0$).
Noting that these modes are already annihilated by $\eth$ and $\eth^\ast$ individually, it is evident from inspecting
eq.~(\ref{eq:EB}) that they are pure gauge modes which do not contribute to the physical
electromagnetic field. Note also that with this operator, the equations of motion would appear to be fourth order
in the angular coordinates. The equations for the true physical degrees of freedom $\eth U$ and $\eth V$, however,
would remain second order in all coordinates.}
\begin{multline}
\label{eq:eom}
\frac{1}{a^3} \partial_t \left(a^3 \partial_t U\right) - \frac{1}{a^2} \Delta U \\+
\left(\frac{\left(\partial_t a\right)^2}{a^2} + \frac{\partial_t^2 a}{a} - \frac{\partial_r^2 s}{a^2 s}\right) U =\Box U + m^2_{\mathrm{eff}} U = 0~,
\end{multline}
and the same for $V$. Since $U$ and $V$ have identical properties, we will from now on only focus on $U$. All results
apply to $V$ in exactly the same way. In the above expression, we have introduced the spatial Laplace operator $\Delta$ for a scalar field  defined as
\begin{equation}
\Delta~=~\frac{1}{s^2} \partial_r \left(s^2 \partial_r\right) + \frac{1}{2s^2} \left( \eth \eth^\ast +\eth^\ast \eth\right)~.
\end{equation}
The operator $\Box$ in (\ref{eq:eom}) is precisely the d'Alembertian for a Lorentz scalar. However, since $U$ is not itself a Lorentz scalar, it also acquires an effective mass $m_{\mathrm{eff}}$ which is precisely the one that corresponds to a conformal coupling to curvature.

In the case of spatial flatness, a complete set of eigenfunctions
to $\Delta$ is given by
\begin{equation}
\label{eq:flatmodes}
X_{p \ell m}(r, \theta, \phi)~=~p \sqrt{2 / \pi} j_\ell (p r) Y_{\ell m}(\theta, \phi)~,
\end{equation}
where we have chosen the spherical Bessel functions $j_\ell$ which are regular at the origin. The eigenvalue equation
on the flat three-dimensional surface is
\begin{equation}
-\Delta X_{p \ell m}~=~p^2 X_{p \ell m}~,
\end{equation}
and the eigenfunctions are normalized as
\begin{equation}
\int\! dr r^2 d\Omega X_{p \ell m} X^\ast_{p'\ell'm'}~=~\delta(p - p') \delta_{\ell\ell'} \delta_{mm'}~.
\end{equation}

In the case of an open geometry where $s = \sinh r$, the eigenfunctions are the harmonics on the three-hyperboloid.
The eigenvalue equation reads
\begin{equation}
-\Delta Y_{p \ell m}~=~\left(p^2 + 1\right) Y_{p \ell m}~,
\end{equation}
where the eigenfunctions $Y_{p \ell m}$ which are regular at $r = 0$ are given by
\begin{multline}
\label{eq:openmodes}
Y_{p \ell m}(r, \theta, \phi)~=~f_{p\ell}(r) Y_{\ell m} (\theta, \phi)~,\\
f_{p\ell}(r) ~\equiv~\frac{\Gamma(i p + \ell + 1)}{\Gamma(i p + 1)} \frac{p}{\sqrt{\sinh r}} P_{i p - 1/2}^{-\ell - 1/2}(\cosh r)~,
\end{multline}
see, e.g., \citet{Sasaki:1994yt}. The normalization is again such that
\begin{equation}
\int\! dr \sinh^2\!r d\Omega Y_{p\ell m} Y^\ast_{p'\ell'm'}~=~\delta(p - p') \delta_{\ell\ell'} \delta_{mm'}~,
\end{equation}
this time on the three-hyperboloid.

The Debye potentials are conformally coupled to gravity, meaning that any conformal factor \textit{which preserves
the spherical symmetry of the geometry} can be absorbed into a redefinition of the fields. In particular this means that
the equation of motion (\ref{eq:eom}) is invariant under a time-dependent conformal
rescaling $g_{\mu\nu} \rightarrow \omega^2(t) g_{\mu\nu}$ (with the corresponding redefinition of time) and a
simple rescaling of the field as $U \rightarrow \omega^{-1}(t) U$.
\vspace{-2pt}


\section{No Supercurvature Modes}
\label{sec:nosupercurv}

\begin{figure}
\centerline{\includegraphics[width=65mm]{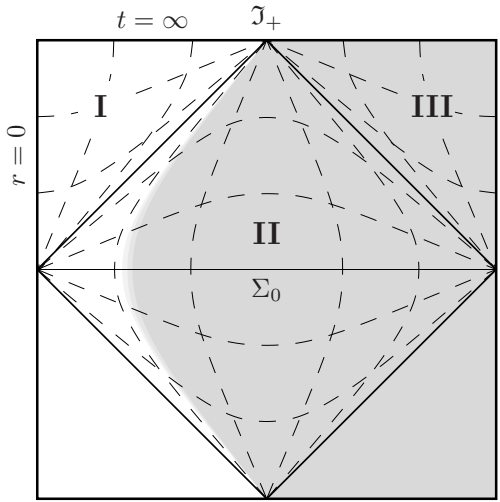}}
\caption{\label{fig:pcd} \small
Spacetime diagram of a one-bubble geometry which may be the result of a Coleman-de~Luccia process.
If one can neglect the geometric effect of the bubble (indicated as the white region),
the spacetime is approximately de~Sitter. There always
exists a conformal map, given as a finite conformal factor, between the $O(3,1)$-symmetric one-bubble spacetime
and the $O(4,1)$-symmetric de~Sitter space. Region I resembles an open FL universe but does not contain any global
Cauchy surface. Such a surface is indicated as $\Sigma_0$, which is entirely contained in region II. Region III is
another open FL universe, similar to, but causally disconnected from region I. Some surfaces
of constant radial or constant time coordinate are indicated as dashed lines.}
\end{figure}

A special situation is given in the open universe models because the three-hyperboloids used in the foliation do not
usually represent global Cauchy surfaces. Therefore, the failure of modes to be normalizable on the hyperboloids does
not necessarily imply that they should be excluded from the physical spectrum. What really matters is the question of whether or not
a mode is normalizable on a Cauchy surface. It is well-known that in certain scalar field models this leads to the
occurrence of modes with discrete imaginary values of $p$ in the spectrum which are usually referred to as
\textit{supercurvature modes}.

In the case of the magnetic field, it was found in \citet{Barrow:2011ic} -- see also references therein to earlier work,
e.g.\ \citet{Barrow:2008jp,Tsagas:2005nn} -- that supercurvature modes, if they exist, give rise to
superadiabatic evolution and can therefore help to solve the problem of magnetogenesis. It is therefore of
relevance whether or not the electromagnetic field can support supercurvature modes. With the formalism of
the Debye potentials, we can now easily address this question.

In order to study whether such supercurvature modes are relevant, we have to check if there are some modes with
imaginary $p$ which are normalizable on a Cauchy
surface. As mentioned before, such a surface can usually not be found within the patch covered by the open coordinate
chart. One therefore has to complete this chart, which means that one usually has to continue the coordinates
across the initial singularity of the open chart (which is a coordinate singularity). As a specific example which
is general enough, one can consider the creation of an open universe by the Coleman-de~Luccia process
\citep{Coleman:1980aw},
as in the open inflation scenario \citep{Bucher:1994gb,GarciaBellido:1997te}. A spacetime diagram is shown in figure~\ref{fig:pcd}. In this type of model,
the open region
(indicated as region I in figure~\ref{fig:pcd}) is contained fully within the lightcone of the nucleation event of a bubble which was created by a vacuum metastability
transition. However, the entire one-bubble spacetime can easily be constructed from the instanton which is responsible
for the transition. A Cauchy surface is then given, e.g., by the  maximal three-section of the instanton, which represents
the so-called turning-point geometry \citep{Coleman:1980aw}. It is located along the horizontal line indicated as $\Sigma_0$ in the figure. Any time-evolution of this surface is, of course, equally suitable.

In order to make the problem tractable analytically, let us ignore for a moment the geometric effects of the bubble
altogether. That is, we consider an exact de~Sitter geometry with $a(t) \equiv \sinh(H t) / H, s(r) \equiv \sinh r$.
For de~Sitter space, the question of supercurvature modes (in a scalar field setting) has been thoroughly studied
in~\citet{Sasaki:1994yt}. It turns out that their analysis can be easily applied to the present setup.
Comparing our eq.~(\ref{eq:eom}) with the equation (2.7) of \citet{Sasaki:1994yt}, it is evident that the
$\nu$-parameter is the one of the conformally coupled field. This should, of course, not come as a surprise since
the electromagnetic field is conformally coupled. The analysis then proceeds with the calculation
of the normalization of the modes on a Cauchy surface. While \citet{Sasaki:1994yt} work in de~Sitter space, we will
present this calculation for an even simpler toy model of an open FL universe, the Milne model, as pedagogical
example in the appendix.

It is shown in \citet{Sasaki:1994yt} that \textit{no supercurvature modes exist} in the conformally coupled case. By
re-applying exactly the same arguments we can therefore conclude that there are no supercurvature modes for the
Debye potentials as well. The result, so far, holds for the case of an exact open de~Sitter background. However, one can
show that it rigorously holds also for any one-bubble geometry like the ones produced in an arbitrary Coleman-de~Luccia
process. This can be seen by noting two facts. Firstly, the radial coordinate on the Cauchy surface corresponds to
the analytic continuation of the time coordinate of the open chart. The continuation of the scale factor $a$ into
the Euclidean domain therefore characterizes the geometric effects of the bubble. It is the behavior in Euclidean time
which determines the normalizability of a mode. Secondly, we note that any bubble geometry can be mapped onto an exact
de~Sitter geometry by a finite conformal factor, which may depend on the radial coordinate on the Cauchy surface.
However, we already pointed out that the mode equation is invariant under such a conformal transformation. In particular,
the normalizability of a mode is not affected by any finite conformal rescaling. In fact this means that any
$O(3, 1)$-symmetric geometry has the same spectrum of modes for the electromagnetic field. The non-existence of
electromagnetic supercurvature modes then follows as a corollary from \citet{Sasaki:1994yt}.

Since the formalism we apply here is very different, it is worthwhile to explain the connection to \citet{Barrow:2011ic}
in some more detail. Given our expression (\ref{eq:EB}) for the magnetic field and using the mode expansion of
eq.~(\ref{eq:openmodes}) for $U$ and $V$, one can show that the covariant three-dimensional
(spatial) vector Laplacian
acting on a magnetic mode with wavenumber $p$ yields
\vspace{-2pt}
\begin{equation}
-\Delta \mathbf{B}_{\left(p\right)}
= \frac{p^2 + 2}{a^2} \mathbf{B}_{\left(p\right)}~.
\end{equation}
A comparison with eq.~($7$) of \citet{Barrow:2011ic} (see also their footnote $6$) then clarifies the relation between our
wavenumber $p$ and their eigenvalue parameterization $n$. The superadiabatic modes with eigenvalues $n^2 < 2$ correspond to
imaginary wavenumbers $p$. We have just shown that these modes are not included in the spectrum in any one-bubble open
universe scenario.
\vspace{-2pt}


\section{Quantum Theory}

A canonical quantization prescription for the Debye potentials works as follows. First, we note that
the true physical degrees of freedom which should be quantized are given by $\eth U$ and $\eth V$. Then, it is advised to rescale the
fields such that the Hubble damping term in the mode equation disappears. To this end, we write $\eth U \equiv \upsilon / a$ and choose a conformal time coordinate defined by $dt = a~d\tau$.
The action for the rescaled field $\upsilon$ reads
\vspace{-2pt}
\begin{multline}
\mathcal{S}_\upsilon~=~\frac{1}{2} \int\!d\tau s^2 dr d\Omega \Bigl[-\partial_\tau \upsilon \partial_\tau \upsilon^\ast + \frac{1}{s^2}
\eth^\ast \upsilon \eth \upsilon^\ast\Bigr.\\ \Bigl. +\frac{1}{s^2} \partial_r \left(s \upsilon\right) \partial_r \left(s
\upsilon^\ast\right)\Bigr]\,.
\end{multline}
Following the usual rules of canonical quantization, the field $\upsilon$ is promoted to an operator $\hat{\upsilon}$
and can be expanded in terms of creation and annihilation operators of modes by writing
\begin{multline}
\hat{\upsilon} (\tau, r, \theta, \phi)~=~\int\!dp \sum_{\ell m} \frac{1}{\sqrt{2}}\Bigl[\hat{\mathrm{a}}_{p \ell m} \upsilon_p (\tau)
{}_{1\!}Y_{p \ell m}(r, \theta, \phi)\Bigr. \\ \Bigl. + \hat{\mathrm{a}}^\dagger_{p \ell m} \upsilon^\ast_p (\tau)
{}^{\phantom\ast\!}_{1\!}Y^\ast_{p \ell m}(r, \theta, \phi)\Bigr]~.
\end{multline}
Note that the field $\upsilon$ is of spin-weight one and should therefore be expanded in terms of the appropriate
spherical harmonics. The eigenfunctions ${}_{1\!}Y_{p \ell m}$ of the spatial Laplace operator are the ones of
eq.~(\ref{eq:openmodes}) with $Y_{\ell m}$ replaced by the corresponding spherical harmonic of spin-weight one,
${}_{1\!}Y_{\ell m}$. In case of spatial flatness, the eigenfunctions ${}_{1\!}Y_{p \ell m}$ have to be replaced by ${}_{1\!}X_{p \ell m}$, which are related to eq.~(\ref{eq:flatmodes}) in a similar way. In both cases, the
mode functions $\upsilon_p(\tau)$ are governed by the mode equation
\begin{equation}
\partial_\tau^2 \upsilon_p +p^2 \upsilon_p~=~0~,
\end{equation}
and it is allowed to choose them as independent of $\ell$ and $m$. If one uses normalized mode functions
\begin{equation}
\mathrm{Im} (\upsilon_p \partial_\tau \upsilon^\ast_p)~=~1~,
\end{equation}
then the equal time commutator of field and canonical momentum,
\begin{multline}
\Bigl[\hat{\upsilon}(\tau, r, \theta, \phi), \partial_\tau \hat{\upsilon}^\ast(\tau, r', \theta', \phi')\Bigr]~=\\ i \frac{1}{s^2}
\delta(r - r') \delta(\cos\theta - \cos\theta') \delta(\phi - \phi')~,
\end{multline}
is equivalent to the standard commutation rules for the creation and annihilation operators,
\begin{equation}
\left[\hat{\mathrm{a}}_{p \ell m}, \hat{\mathrm{a}}^\dagger_{p'\ell'm'}\right]~=~\delta(p - p') \delta_{\ell\ell'} \delta_{mm'}~.
\end{equation}

The reader may wonder what would have been the difference if one had quantized the scalar Debye
potentials directly instead of the helicity-one degrees of freedom which we obtained by applying a spin-raising
operator. Firstly, by choosing to quantize the latter, we have avoided a quantization of the unphysical gauge modes
with $\ell = 0$ which are present in the expansion of a scalar but not in the one of the helicity-one fields.
Secondly, by noting that $\eth Y_{\ell m} = \sqrt{\ell \left(\ell + 1\right)} {}_{1\!}Y_{\ell m}$, one can see
that some $\ell$-dependent factors may appear in the commutation rules for the scalar Debye potentials. A careful
look at the action reveals that the scalar modes are not canonically normalized and that these factors
are therefore expected. These differences, however, are completely inessential for the question of whether
or not any supercurvature modes are part of the spectrum. In particular, the argument of section~\ref{sec:nosupercurv}
works equally well for the helicity-one degrees of freedom, with rather obvious modifications when going through the
detailed proof.

The standard choice of positive frequency mode functions is like in Minkowski space,
\begin{equation}
\upsilon_p(\tau)~=~\frac{1}{\sqrt{p}} e^{-i p \tau}~,\qquad\qquad\text{(Minkowski)} \,.
\end{equation}
This is not surprising as the rescaled electromagnetic fields, $B/a^2$ and $E/a^2$ are independent of
the scale factor in conformal time. One can verify that
some standard results of quantized electromagnetism are reproduced. We checked this for the vacuum two-point correlators
$\langle E_a(t, r, \theta, \phi) E_b(t', r', \theta', \phi')\rangle$, which turn out to be the same as if obtained from a
standard quantization of $A^\mu$ in Minkowski space.

However, the above quantization prescription is general enough to be applicable also in arbitrary flat or open FL backgrounds.
For instance, one could obtain the primordial power spectrum of the electromagnetic field in the open inflation scenario.
Since we have proven that no supercurvature modes
are present, the result of this computation is, however, of pure academic interest, because one does not expect that significant
perturbation amplitudes can be obtained after the inflationary era.
\vspace{-2pt}

\section{Conclusions}

In this letter we have studied the modes of the quantum vacuum of the electromagnetic field
in an open, inflating Friedmann universe. Whilst subcurvature modes decay extremely fast after their generation,  supercurvature modes, on the other hand, could remain relevant at the end of inflation and could then have a significant impact on the origin of large scale magnetic fields in the present Universe. Great care should therefore be taken in understanding under which circumstances supercurvature modes are expected to belong to the magnetic field spectrum.  Here we have explored the eigenmodes of the electromagnetic field
in an open universe and we have shown that supercurvature modes are not expected to be produced via any causal process, such as if the open universe is generated by bubble nucleation. This is a consequence of the conformal coupling of electromagnetism.

If one switches on some perturbative interaction later during inflation, conformal invariance may be broken and electromagnetic modes may be generated. But only modes which are present in the quantum vacuum
can be excited by such a perturbative coupling, namely subcurvature modes.

We have focused on the open inflation scenario with the remark that it presently is
the only physically motivated scenario which procures an open Friedmann universe and offers a complete
enough framework to address the question of supercurvature modes. It also represents the
scenario preferred by recent considerations of eternal inflation and the Landscape of string theory.
Our argument shows already that no supercurvature modes
can exist in all cases where the global spacetime carries the same $O(3, 1)$-symmetry as is displayed by the open
patch. In less specific settings which are deprived of a global
description, the question may be elusive, but the experience with the open inflation scenario teaches us that the
physical viability of supercurvature modes in the context of standard electromagnetism has yet to be demonstrated.
In a singular open Friedmann universe this question cannot be seriously addressed, and when addressed naively, choosing $\{t=\rm{const.}\}$ hypersurfaces, again no supercurvature modes are contained in the physical spectrum.

In view of this result, it appears more and more unlikely that standard electromagnetism can support
any superadiabatic evolution of cosmological magnetic fields under physical conditions. Such an evolution
can be obtained only by breaking the conformal invariance of electromagnetism (already at the time of bubble nucleation)
or by introducing some other kind of new physics.
\vspace{-2pt}


\section*{\small Acknowledgments}
We thank Misao Sasaki for interesting comments.
JA wants to thank the University of Geneva for hospitality and the German Research Foundation (DFG)
for financial support through the Research Training Group 1147 ``Theoretical Astrophysics and
Particle Physics.'' CdR and RD are supported by the Swiss National Science Foundation.
\vspace{-2pt}

\appendix

\section{Mode Spectrum in the Milne Universe}

We present here the explicit computation of the mode spectrum in the Milne model, which is one of the simplest
open FL geometries. It is obtained by rewriting the line element of Minkowski space as
\begin{equation}
ds^2 = -dT^2 + dR^2 + R^2 d\Omega^2 = -dt^2 + t^2 \left[dr^2 + \sinh^2 r d\Omega^2\right]~.
\end{equation}
While $-\infty < T < \infty$ and $0 \leq R < \infty$ are the coordinates of a standard (spatially flat)
spherical coordinate system which covers the full Minkowski spacetime, using $0 < t < \infty$ and $0 \leq r < \infty$
one obtains a metric of the open FL type with $a = t$, cf.\ eq.~(\ref{eq:metric}). This new coordinate system,
with $T=t\cosh r$ and $R=t\sinh r$,
covers the interior of the future lightcone of $T = R = 0$. Within this very simple setting which yet has all the
desired features, we want now to exemplify the reasoning of section \ref{sec:nosupercurv}.

As a first step, for the mode expansion of eq.~(\ref{eq:openmodes}) we can immediately solve the mode equation.
The solutions to eq.~(\ref{eq:eom}) take the form
\begin{equation}
U_{p \ell m \pm}(t, r, \theta, \phi) = N_{p\pm} t^{-1 \pm i p} Y_{p \ell m}(r, \theta, \phi)~,
\end{equation}
where $N_{p\pm}$ is a normalization to be determined. To this end, we want to evaluate the Klein-Gordon inner product
on a Cauchy surface. The whole point is that the open spatial hypersurfaces $\lbrace t = \mathrm{const.}\rbrace$ do not represent
proper Cauchy surfaces and one should therefore make a better choice. We choose the surface $\lbrace T = 0\rbrace$
which is a proper global section of Minkowski space. This hypersurface lies entirely outside the coordinate patch
covered by $t, r$, however, by making appropriate analytic continuations, we can complete the chart to
 include $\lbrace T = 0\rbrace$.
More precisely, by taking $t \rightarrow i \rho$, $r \rightarrow \tau - i \pi / 2$, the region outside the lightcone
is covered by $-\infty < \tau < \infty$ and $0 < \rho < \infty$. Furthermore, the hypersurface
$\lbrace T = 0\rbrace$ coincides with the one defined by $\lbrace \tau = 0\rbrace$. The line element is given as
\begin{equation}
ds^2 = d\rho^2 - \rho^2 d\tau^2 + \rho^2 \cosh^2\tau d\Omega^2~.
\end{equation}

It is noteworthy that the role of time and radial distance have been interchanged by the analytic continuation,
just as it was done by \citet{Sasaki:1994yt} for the case of de~Sitter.
The Klein-Gordon inner product is finally given by
\begin{multline}
\label{eq:KGproduct}
\langle U_{p \ell m \pm}, U_{p' \ell' m' \pm} \rangle_{\mathrm{K-G}} = i\! \int\limits_{T = 0}\!dR R^2 d\Omega U_{p \ell m \pm}^\ast \overset{\leftrightarrow}{\partial_T} U_{p' \ell' m' \pm}\\
= i \delta_{\ell\ell'} \delta_{mm'} N^\ast_{p\pm} N_{p'\pm} e^{\mp\left(p + p'\right) \pi / 2} \int\limits_0^\infty \frac{d\rho}{\rho} \rho^{\mp i \left(p - p'\right)} \\\times \cosh^2 \tau \left. f_{p\ell}^\ast \overset{\leftrightarrow}{\partial_\tau} f_{p'\ell'} \right|_{\tau = 0}
\end{multline}
By making a change of variables to $\ln \rho$, one can see that the $\rho-$integral is a representation of the delta
function $\delta(p - p')$ for $p, p'$ real. For any imaginary $p$ or $p'$, the integral is badly divergent, which
implies that the modes with imaginary $p$ have zero norm. In other words, there are no supercurvature
modes of the Debye potentials in the Milne model.

For the regular modes with real values of $p$, the term printed in the last line of eq.~(\ref{eq:KGproduct}) can be
easily evaluated once setting $p = p'$ and $\ell = \ell'$. One obtains
\begin{equation}
\cosh^2 \tau \left. f_{p\ell}^\ast \overset{\leftrightarrow}{\partial_\tau} f_{p\ell} \right|_{\tau = 0} = -\frac{2 i p}{\pi} \sinh \pi p~.
\end{equation}

\bibliographystyle{mn2e}
\bibliography{supercurv}

\begin{thebibliography}{}

\bibitem[\protect\citeauthoryear{Banerjee \& Jedamzik}{Banerjee \&
  Jedamzik}{2004}]{Banerjee:2004df}
Banerjee R.,  Jedamzik K.,  2004, Phys. Rev., D70, 123003

\bibitem[\protect\citeauthoryear{Barrow \& Tsagas}{Barrow \&
  Tsagas}{2008}]{Barrow:2008jp}
Barrow J.~D.,  Tsagas C.~G.,  2008, Phys. Rev., D77, 107302

\bibitem[\protect\citeauthoryear{Barrow \& Tsagas}{Barrow \&
  Tsagas}{2011}]{Barrow:2011ic}
Barrow J.~D.,  Tsagas C.~G.,  2011, Mon. Not. Roy. Astron. Soc., 414, 512

\bibitem[\protect\citeauthoryear{Battaglia, Pfrommer, Sievers, Bond \&
  Ensslin}{Battaglia et~al.}{2009}]{Battaglia:2008ex}
Battaglia N.,  Pfrommer C.,  Sievers J.~L.,  Bond J.~R.,    Ensslin T.~A.,
  2009, Mon. Not. Roy. Astron. Soc., 393, 1073

\bibitem[\protect\citeauthoryear{Bucher, Goldhaber \& Turok}{Bucher
  et~al.}{1995}]{Bucher:1994gb}
Bucher M.,  Goldhaber A.~S.,    Turok N.,  1995, Phys. Rev., D52, 3314

\bibitem[\protect\citeauthoryear{Campanelli}{Campanelli}{2007}]{Campanelli:200%
7tc}
Campanelli L.,  2007, Phys. Rev. Lett., 98, 251302

\bibitem[\protect\citeauthoryear{Caprini, Durrer \& Fenu}{Caprini
  et~al.}{2009}]{Caprini:2009pr}
Caprini C.,  Durrer R.,    Fenu E.,  2009, JCAP, 0911, 001

\bibitem[\protect\citeauthoryear{Clarke, Kronberg \& B{\"o}hringer}{Clarke
  et~al.}{2001}]{Clarke:2000bz}
Clarke T.~E.,  Kronberg P.~P.,    B{\"o}hringer H.,  2001, Astrophys. J., 547,
  L111

\bibitem[\protect\citeauthoryear{Coleman \& De~Luccia}{Coleman \&
  De~Luccia}{1980}]{Coleman:1980aw}
Coleman S.~R.,  De~Luccia F.,  1980, Phys. Rev., D21, 3305

\bibitem[\protect\citeauthoryear{De~Simone \& Salem}{De~Simone \&
  Salem}{2010}]{DeSimone:2009dq}
De~Simone A.,  Salem M.~P.,  2010, Phys. Rev., D81, 083527

\bibitem[\protect\citeauthoryear{Durrer, Hollenstein \& Jain}{Durrer
  et~al.}{2011}]{Durrer:2010mq}
Durrer R.,  Hollenstein L.,    Jain R.~K.,  2011, JCAP, 1103, 037

\bibitem[\protect\citeauthoryear{Garc{\'i}a-Bellido, Garriga \&
  Montes}{Garc{\'i}a-Bellido et~al.}{1998}]{GarciaBellido:1997te}
Garc{\'i}a-Bellido J.,  Garriga J.,    Montes X.,  1998, Phys. Rev., D57, 4669

\bibitem[\protect\citeauthoryear{Garc{\'i}a-Bellido, Liddle, Lyth \&
  Wands}{Garc{\'i}a-Bellido et~al.}{1995}]{GarciaBellido:1995wz}
Garc{\'i}a-Bellido J.,  Liddle A.~R.,  Lyth D.~H.,    Wands D.,  1995, Phys.
  Rev., D52, 6750

\bibitem[\protect\citeauthoryear{Goldberg, MacFarlane, Newman, Rohrlich \&
  Sudarshan}{Goldberg et~al.}{1967}]{Goldberg:1966uu}
Goldberg J.,  MacFarlane A.,  Newman E.,  Rohrlich F.,    Sudarshan E.,  1967,
  J. Math. Phys., 8, 2155

\bibitem[\protect\citeauthoryear{Guth}{Guth}{2007}]{Guth:2007ng}
Guth A.~H.,  2007, J. Phys., A40, 6811

\bibitem[\protect\citeauthoryear{Jaffe et~al.,}{Jaffe
  et~al.}{2001}]{Jaffe:2000tx}
Jaffe A.~H.,  et~al., 2001, Phys. Rev. Lett., 86, 3475

\bibitem[\protect\citeauthoryear{Kronberg}{Kronberg}{1994}]{Kronberg:1993vk}
Kronberg P.~P.,  1994, Rept. Prog. Phys., 57, 325

\bibitem[\protect\citeauthoryear{Linde}{Linde}{1986}]{Linde:1986fd}
Linde A.~D.,  1986, Phys. Lett., B175, 395

\bibitem[\protect\citeauthoryear{Lyth \& Woszczyna}{Lyth \&
  Woszczyna}{1995}]{Lyth:1995cw}
Lyth D.~H.,  Woszczyna A.,  1995, Phys. Rev., D52, 3338

\bibitem[\protect\citeauthoryear{Mo \& Papas}{Mo \&
  Papas}{1972}]{TseChin:1973mp}
Mo T.~C.,  Papas C.~H.,  1972, Phys. Rev., D6, 2071

\bibitem[\protect\citeauthoryear{{Neronov} \& {Vovk}}{{Neronov} \&
  {Vovk}}{2010}]{Neronov:2010}
{Neronov} A.,  {Vovk} I.,  2010, Science, 328, 73

\bibitem[\protect\citeauthoryear{Pentericci, Van~Reeven, Carilli, Rottgering \&
  Miley}{Pentericci et~al.}{2000}]{Pentericci:2000mp}
Pentericci L.,  Van~Reeven W.,  Carilli C.~L.,  Rottgering H. J.~A.,    Miley
  G.~K.,  2000, Astron. Astrophys. Suppl. Ser., 145, 121

\bibitem[\protect\citeauthoryear{Sasaki, Tanaka \& Yamamoto}{Sasaki
  et~al.}{1995}]{Sasaki:1994yt}
Sasaki M.,  Tanaka T.,    Yamamoto K.,  1995, Phys. Rev., D51, 2979

\bibitem[\protect\citeauthoryear{Spergel et~al.,}{Spergel
  et~al.}{2003}]{Spergel:2003cb}
Spergel D.~N.,  et~al., 2003, Astrophys. J. Suppl., 148, 175

\bibitem[\protect\citeauthoryear{Susskind}{Susskind}{2007}]{CarrSusskind}
Susskind L.,  2007, in Carr B.,  ed., {Universe or Multiverse?} {The Anthropic
  Landscape of String Theory}.
Cambridge Univ. Press, pp 247--266

\bibitem[\protect\citeauthoryear{Taylor, Vovk \& Neronov}{Taylor
  et~al.}{2011}]{Taylor:2011bn}
Taylor A.,  Vovk I.,    Neronov A.,  2011, Astron. Astrophys., 529, A144

\bibitem[\protect\citeauthoryear{Tsagas \& Kandus}{Tsagas \&
  Kandus}{2005}]{Tsagas:2005nn}
Tsagas C.~G.,  Kandus A.,  2005, Phys. Rev., D71, 123506

\bibitem[\protect\citeauthoryear{Weinberg}{Weinberg}{2007}]{CarrWeinberg}
Weinberg S.,  2007, in Carr B.,  ed., {Universe or Multiverse?} {Living in the
  Multiverse}.
Cambridge Univ. Press, pp 29--42

\end{thebibliography}

\end{document}